\documentclass[letterpaper,11pt]{article}

\usepackage{fancyhdr}
\usepackage{amsmath}
\usepackage{amsbsy}
\usepackage{amssymb}
\usepackage{amsthm}
\usepackage{amscd}
\usepackage{amsfonts}
\usepackage{supertabular}
\usepackage{graphics}
\usepackage{graphicx}
\usepackage{verbatim}
\usepackage{subfigure}
\usepackage{epsfig}
\usepackage{xspace}
\usepackage{euscript}
\usepackage{alltt}
\usepackage{boxedminipage}
\usepackage{float}
\usepackage{times}
\usepackage[colorlinks]{hyperref}
\usepackage[square,numbers]{natbib}
\usepackage{setspace}
\usepackage{amsmath, amssymb}
\usepackage{algorithm}
\usepackage{booktabs}
\usepackage{algorithmic}
\usepackage{lscape}
\usepackage{wrapfig}
\usepackage{subfigure}%
\usepackage[hang,small,bf]{caption}
\usepackage{color}
\usepackage{authblk}
\usepackage[pagewise,displaymath]{lineno}

\def\Put(#1,#2)#3{\leavevmode\makebox(0,0){\put(#1,#2){#3}}}

\begin{document}

\nolinenumbers

\title{General and mechanistic optimal relationships for tensile strength of doubly convex tablets under diametrical compression}

\author[a]{Sonia M. Razavi}
\author[b]{Marcial Gonzalez \thanks{Corresponding author at: School of Mechanical Engineering, Purdue University, West Lafayette, IN 47907, USA. Tel.:~+1~765 494 0904. Fax: +1 765 496 7537 \\ \indent E-mail address: marcial-gonzalez@purdue.edu (M. Gonzalez)}}
\author[a]{Alberto M. Cuiti\~{n}o}
\affil[a]{\small Department of Mechanical and Aerospace Engineering, Rutgers University, Piscataway, NJ 08854, USA}
\affil[b]{\small School of Mechanical Engineering, Purdue University, West Lafayette, IN 47907, USA}

\maketitle

\begin{abstract}
We propose a general framework for determining optimal relationships for tensile strength of doubly convex tablets under diametrical compression. This approach is based on the observation that tensile strength is directly proportional to the breaking force and inversely proportional to a non-linear function of geometric parameters and materials properties. This generalization reduces to the analytical expression commonly used for flat faced tablets, i.e., Hertz solution, and to the empirical relationship currently used in the pharmaceutical industry for convex-faced tablets, i.e., Pitt's equation. Under proper parametrization, optimal tensile strength relationship can be determined from experimental results by minimizing a figure of merit of choice. This optimization is  performed under the first-order approximation that a flat faced tablet and a doubly curved tablet have the same tensile strength if they have the same relative density and are made of the same powder, under equivalent manufacturing conditions. Furthermore, we provide a set of recommendations and best practices for assessing the performance of optimal tensile strength relationships in general. Based on these guidelines, we identify two new models, namely the \emph{general and mechanistic models}, which are effective and predictive alternatives to the tensile strength relationship currently used in the pharmaceutical industry.
\end{abstract}


\section{Introduction}

\begin{figure}[htb]
\centering
\subfigure[]
{
\includegraphics[scale=0.60]{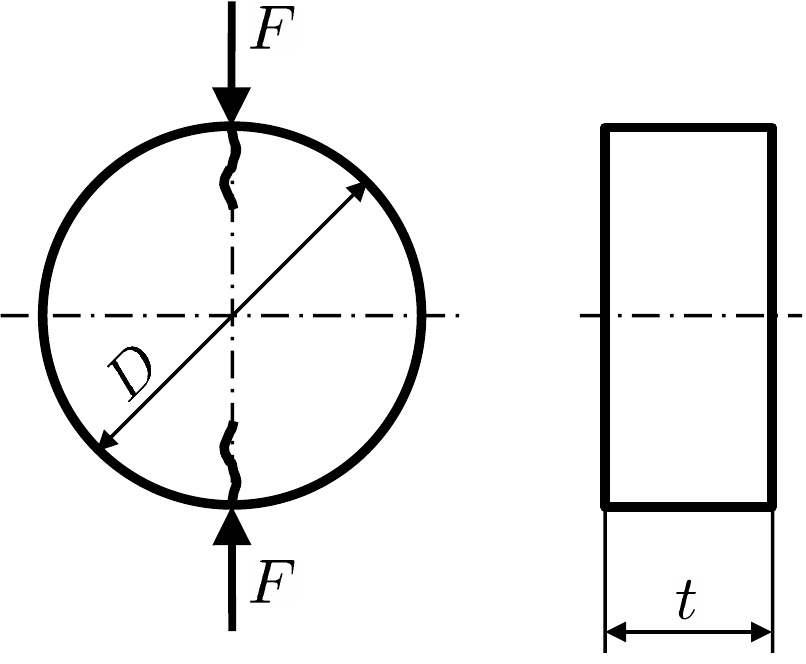}
\label{Fig:Tablet-shape}
}
\hspace{.8in}
\subfigure[]
{
\includegraphics[scale=0.60]{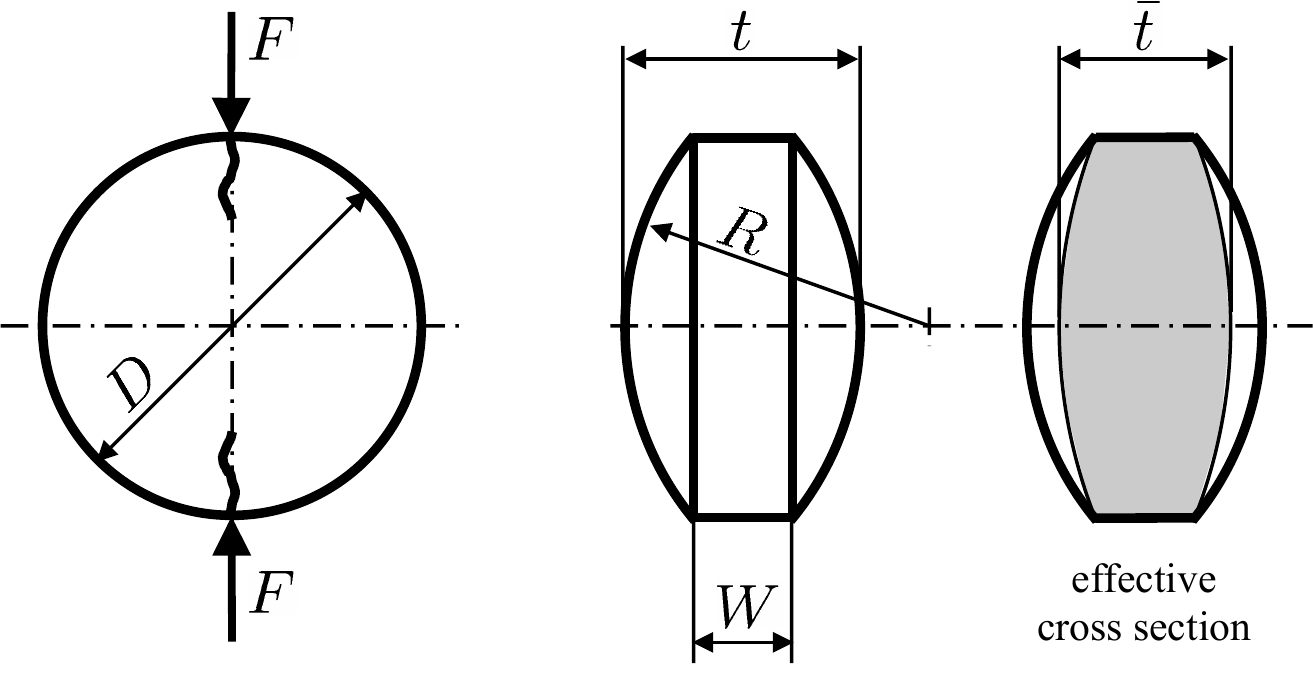}
\label{Fig:Tablet-failure}
}
\caption{Geometry and failure behavior (a) in a flat faced cylindrical tablet and (b) in a doubly convex tablet under diametrical compression.}
\end{figure}

Pharmaceutical tablets are fabricated by pressing powders into various shapes and geometries. It is important to assure that tablets have sufficient strength to endure post-compaction loading such as coating, packaging, handling and storage. Tablet strength is thus an important quality factor that is tested during tablet production \cite{Pharmacopeia-2011}. There have been many efforts to establish destructive and non-destructive techniques to determine the mechanical strength of compacted powders (see, e.g.,  \cite{Podczeck-2012} and references therein). Among these experimental techniques, the diametrical compression test, also referred to as Brazilian test \cite{Barcellos-1953}, is the most conventional method used in the pharmaceutical industry to measure the breaking force of a tablet. The test consists in placing and compressing a tablet along its diameter between two rigid platens. Under the assumption of linear elastic behavior  \cite{Timoshenko-1970}, the breaking force of a cylindrical tablet can be related to its tensile strength by the following expression, known as Hertz solution,
 \begin{equation}
 \sigma_t = \frac{2F}{\pi D t}
\label{Eq:Hertz}
\end{equation}
where $\sigma_t$ is the tensile strength, $F$ is the breaking force, $D$ is the diameter of the tablet, and $t$ is its thickness as shown in Fig.~\ref{Fig:Tablet-shape}. The above expression is only valid for flat cylindrical tablets that fail in tension across the symmetry plane of the loaded diameter. 

Tensile strength and breaking force increase exponentially with increasing relative density for typical pharmaceutical powders, tableting speeds and tablet shapes (see, e.g., \cite{Tye-2005,Sinka-2009,Haririan-1999}). In addition, the breaking force exhibits a strong dependence on the shape of the tablet and only a mild dependence on the compaction speed. Unfortunately, and in sharp contrast to flat faced tablets, there is no closed-form analytical solution that relates tensile strength and breaking force for curved faced tablets (cf. Eq.~\ref{Eq:Hertz}). In order to amend this situation, Pitt \emph{et al.} \cite{Pitt-1989} used a photoelastic method to measure the stress distribution of doubly convex tablets subject to diametrical load. For doubly convex cylindrical gypsum discs, they established the following empirical relationship \cite{Pitt-1988} between geometric parameters, breaking force and tensile strength
\begin{equation}
\label{Eq:Pitt}
\sigma_t = \frac{10F}{\pi D^2 \left[2.84 \left(\frac{t}{D}\right) - 0.126 \left(\frac{t}{W}\right) +3.15\left(\frac{W}{D}\right)+ 0.01\right]}
\end{equation}
where $W$ is the length of the cylindrical portion of the tablet (see Fig.~\ref{Fig:Tablet-failure}). This equation is valid for any brittle doubly convex disc with $0.1\le W/D \le0.3$, and also for discs with $W/D=0.06$ and $D/R<1.0$. However, it is worth noting that Pitt's equation \eqref{Eq:Pitt} does not reduce to the Hertz solution \eqref{Eq:Hertz} when the geometric parameters correspond to those of a flat faced tablet. Pitt and co-workers \cite{Pitt-2013} subsequently modified Eqs.~\eqref{Eq:Hertz} and \eqref{Eq:Pitt} to be applicable for elongated tablets by multiplying both equations by a factor of $2/3$---this factor is exact only for the limiting case of large length to width ratios.

Shang \emph{et al.}  \cite{Shang-2013a} adopted the form of Pitt's original equation and calibrated its empirical coefficients to an extensive experimental campaign of doubly convex microcrystalline cellulose tablets with various curvatures and of various relative densities. Specifically, they fit experimental measurements to
 \begin{equation}
 \sigma_t = \frac{F}{\pi D^2 \left[ a\left(\frac{t}{D}\right)+b\left(\frac{t}{W}\right)+c\left(\frac{W}{D}\right)+d\right]}
\label{Eq:4-parameter}
\end{equation}
where $a$, $b$, $c$, and $d$ are empirical coefficients (see Table~{\ref{Table-MaterialProperties-I} for numerical values). This equation, which from now on is referred to as the \emph{4-parameter model}, has the same application space as Pitt's equation but shows a better fit to experimental data. They also simplified \eqref{Eq:4-parameter} by observing that there are only two independent geometric parameters (e.g., $t/D$ and $W/D$) and that the correct limiting behavior for flat geometries can be enforced analytically. Thus, they proposed
 \begin{equation}
 \sigma_t = \frac{F}{\pi D^2 \left[a\left(\frac{t}{D}\right)+c\left(\frac{W}{D}\right)\right]}
\label{Eq:Shang}
\end{equation}
which we refer to as the \emph{1-parameter model}, where $a=0.14$ and $c=0.5-a=0.36$ are empirical parameters. It is interesting to note that Shang and co-workers reported in \cite{Shang-2013b} that the optimal values for $a$ and $c$ do not necessarily sum up to one-half (i.e., they do not enforce the correct limiting behavior in \eqref{Eq:Shang}) when calibrated to detailed finite element numerical results (e.g., $a=0.187$ and $c=0.284$ are proposed). Furthermore, and in contrast to Shang's results, Podczeck \emph{et al.} \cite{Podczeck-2013} calibrated finite element simulations to $a=0$ and $c=0.5$, for doubly convex geometries with $0.06\le W/D \le0.5$ and $D/R \le1.85$ which fail in accord with Fig.~\ref{Fig:Tablet-failure}. These results suggest that the elucidations of optimal tensile strength relationships and of optimal procedures to calibrate their parameters are important areas worthy of further research. 

In the present work we propose a general framework for determining optimal relationships for tensile strength of doubly convex tablets under diametrical compression. This approach is based on the observation that tensile strength is directly proportional to the breaking force and inversely proportional to a non-linear function of geometric parameters and materials properties. Under proper parametrization, the tensile strength relationship can be determined from experimental results by solving an optimization problem that minimizes a figure of merit of choice. Based on this general framework, we develop three new optimal tensile strength relationships and three different figures of merit to determine their optimal parameters. We also provide a set of guidelines for assessing the performance of optimal tensile strength relationships, with which we compare the new models with two models previously proposed in the literature (i.e., equations \eqref{Eq:4-parameter} and \eqref{Eq:Shang}). This analysis reveals that two of the new models, namely the general and mechanistic models, are effective and predictive alternatives to the tensile strength relationship currently used in the pharmaceutical industry.

\section{Optimal tensile strength relationships}

The tensile strength $\sigma_t$ is related to the breaking force $F$ under diametrical compression by the following general equation 
\begin{equation}
\sigma_t = \frac{F}{\pi D^2 Q}
\label{Eq:General-Q}
\end{equation}
where $Q$ is a nonlinear function of geometric parameters and material properties. For example, for flat faced elastic isotropic cylindrical tablets, $Q=t/2D$ when the tablet is under concentrated loads \cite{Timoshenko-1970} and $Q=t/2D[1-(b/D)^2]^{3/2}$ when the tablet is under loads uniformly distributed on a stripe of width $b$ \cite{Tang-1994}. Analytical expressions for $Q$ can also be derived under the assumption of radial pressures acting on the tablet (see, e.g., \cite{Hondros-1959} for uniform radial pressure and \cite{Kourkoulis-2012} for parabolic radial pressure) and for flattened cylinders subject to uniform diametrical compression \cite{Wang-2004}. The function $Q$ may additionally account for the effect of anisotropy \cite{Bagault-2012}, plastic behavior \cite{Procopio-2003}, and lack of plane stress conditions, among other material and geometric characteristics. 

Here we assume that $\sigma_t$ is a material property that solely depends on the relative density $\rho_R$ of the tablet, for given powder and manufacturing conditions. The function $\sigma_t(\rho_R)$ can then be readily obtained from an experimental campaign of flat faced cylindrical tablets.  We also assume that $Q$ is a geometric function that does not depend on powder properties or manufacturing variables. Based on such assumptions, the function $Q$ can be determined from experimental results by solving the following optimization problem
\begin{linenomath*}
\begin{eqnarray*}
	\min_{Q: \mathcal{G} \rightarrow \mathbb{R}} 
		\left[ 
			\sum_{\{G_i,\rho_i\} \in \mathcal{G}\times\mathcal{D}}  
			\left( \sigma_t(\rho_i) - \frac{F_i}{\pi~D_i^2~Q(G_i)} \right)^2 
		\right]^{1/2} 
	=:
	\min_{Q: \mathcal{G} \rightarrow \mathbb{R}}  \sigma\mathrm{-norm}
\end{eqnarray*}
\end{linenomath*}
where $\mathcal{G}$ is the space of all possible tablet geometries, $\mathcal{D}$ is an interval of tablet relative densities (e.g., from the critical or jamming density to full compaction or relative density of 1.00), and $D_i$ is the diameter of geometry $G_i$ along which the breaking force $F_i$ is applied. In the above expression, $\sigma_t(\rho_i)$ is obtained from a flat faced cylindrical made with the same powder and under the same manufacturing conditions employed for making the tablet with geometry $G_i$ and relative density $\rho_i$. Alternatively, the function $Q$ can be determined by solving any of the following equivalent problems
\begin{linenomath*}
\begin{eqnarray*}
	\min_{Q: \mathcal{G} \rightarrow \mathbb{R}} 
		\left[ 
			\sum_{\{G_i,\rho_i\} \in \mathcal{G}\times\mathcal{D}}  
			\left( Q(G_i) - \frac{F_i}{\pi~D_i^2~\sigma_t(\rho_i)} \right)^2 
		\right]^{1/2} 
	=:
	\min_{Q: \mathcal{G} \rightarrow \mathbb{R}}  Q\mathrm{-norm}
\end{eqnarray*}
\end{linenomath*}
\begin{linenomath*}
\begin{eqnarray*}
	\min_{Q: \mathcal{G} \rightarrow \mathbb{R}} 
		\left[ 
			\sum_{\{G_i,\rho_i\} \in \mathcal{G}\times\mathcal{D}}  
			\left( Q(G_i)~\sigma_t(\rho_i) - \frac{F_i}{\pi~D_i^2} \right)^2 
		\right]^{1/2} 
	=:
	\min_{Q: \mathcal{G} \rightarrow \mathbb{R}}  Q\sigma\mathrm{-norm}
\end{eqnarray*}
\end{linenomath*}

It bears emphasis that these equivalent optimization problems determine $Q$ by enforcing Eq.~\eqref{Eq:General-Q}, that is they determine an optimal tensile strength relationship. If there were no experimental uncertainty, these three optimization problems behave similarly and have the same solution. However, in reality, experimental uncertainty is unavoidable and the form of $Q$ has to be approximated and parametrized. Thus, our goal is to find the optimal form for $Q$ and the most stable optimization problem to determine its fitting parameters (i.e., for example, the minimization of the $Q$-norm, the $\sigma$-norm or the $Q\sigma$-norm). 

In the interest of applicability, we restrict attention to doubly convex tablets whose geometries can be parametrized by $t/D$, $t/W$, $W/D$ as depicted in Fig.~\ref{Fig:Tablet-failure}. Specifically, we consider microcrystalline cellulose tablets for which Shang \emph{et al.} \cite{Shang-2013a} have obtained the relationship between tensile strength and relative density using flat faced tablets (see Fig.~\ref{Fig: stress-RD} and Table~\ref{tab: stress-RD} in the Appendix for the numerical values extracted from \cite{Shang-2013a}). The experimental data is best fit to an exponential function, that is
 \begin{equation}
 \sigma_t = A~\mathrm{e}^{B\rho_R}
\label{eq:RD}
\end{equation}
where $A=24.68$ kPa and $B=6.516$. Shang and co-workers have additionally reported results for an extensive experimental campaign of doubly convex tablets with various curvatures and of various relative densities. The relative density of each tablet is computed by dividing the tablet density over the material true density, i.e., $1590$ kg/m$^3$ for microcrystalline cellulose \cite{Shang-2013b}.

\begin{figure}[htb]
\centering
\includegraphics[scale=0.41]{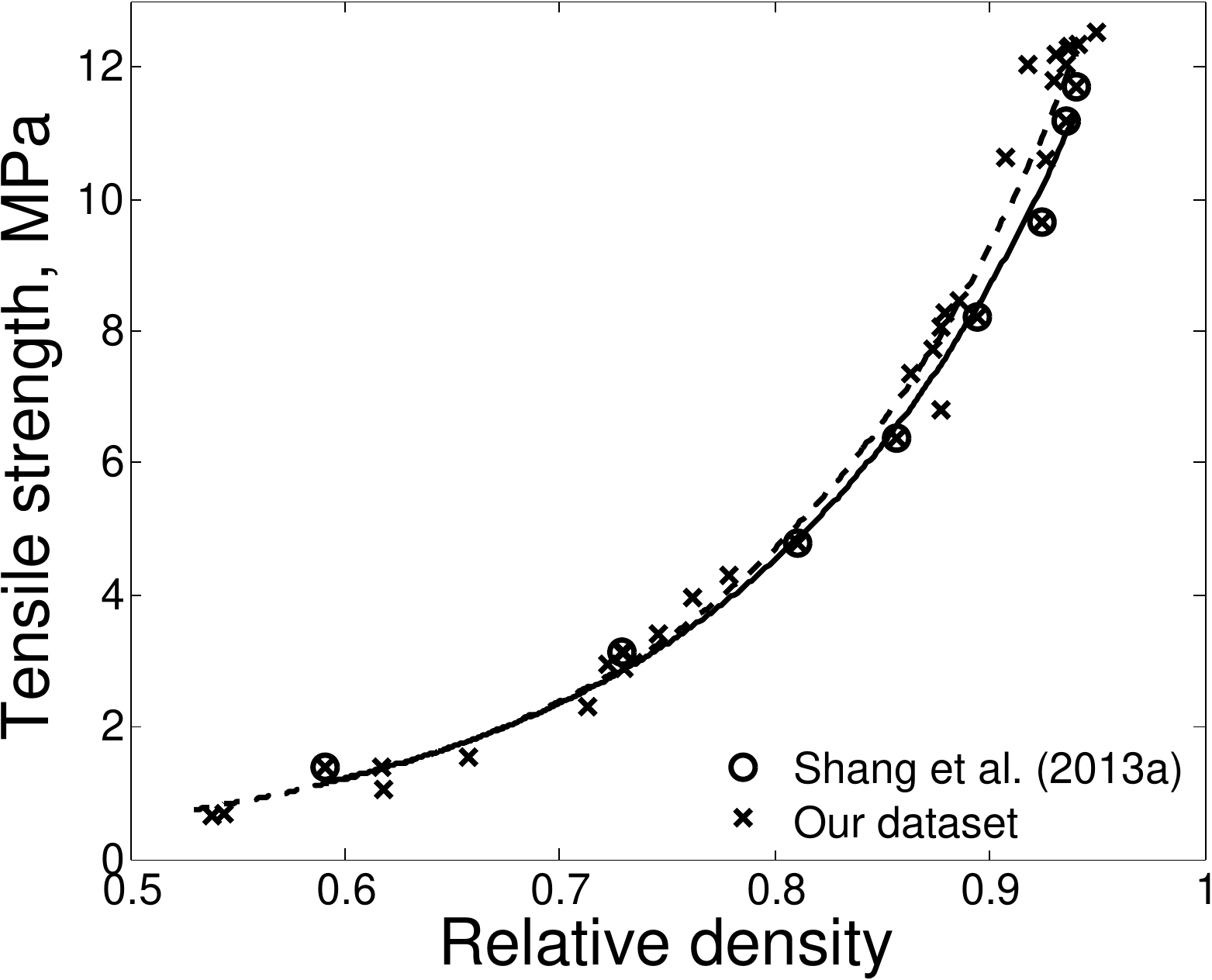}
\caption{Relationship between tensile strength and relative density of flat faced tablets. According to Eq.(\ref{eq:RD}), $A$ and $B$ are 24.68 kPa and 6.516 for the full curve and 20.65 kPa and 6.787 for the dashed curve, respectively.}
\label{Fig: stress-RD}
\end{figure}

For the sake of simplicity, we assume a form for $Q$ that only captures the leading order term of each geometric parameter, that is
\begin{linenomath*}
\begin{eqnarray*}
	Q = a \left(\frac{t}{D}\right)^e + b \left(\frac{t}{W}\right)^f + c \left(\frac{W}{D}\right)^g + d
\end{eqnarray*}
\end{linenomath*}
with the constraint that $Q \rightarrow t/2D$ as $W \rightarrow t$, in order to enforce the correct limit for flat tablets. Thus, the relationship between tensile strength, breaking force and geometric parameters, i.e., Eq.~\eqref{Eq:General-Q}, simplifies to
\begin{equation}
    \sigma_t 
    = 
    \frac{F}
         {\pi D^2 \left[ a \left(\frac{t}{D}\right)^e + b \left(\frac{t}{W}\right)^f + c \left(\frac{W}{D}\right)^g + d \right]}
\label{eq:general}
\end{equation}
where the parameters $\{a,b,c,d,e,f,g\}$ are either assumed known or optimally estimated from a set of experimental observations (e.g., Shang's dataset \cite{Shang-2013a}) using different figures of merit (e.g., $\sigma$-norm, $Q$-norm, and $Q\sigma$-norm). If $e=f=g=1$, the 4-parameter model is recovered, i.e., Eq.~\eqref{Eq:4-parameter}. However, if Pitt's coefficients are used in the 4-parameter model then the correct limit for flat tablets is not attained, indicating that there is a different set of coefficients that further minimize the problem presented above. Similarly, if $b=0$, $c=1/2-a$ and $e=g=1$, the 1-parameter model is recovered, i.e., Eq.~\eqref{Eq:Shang}. Table~\ref{Table-MaterialProperties-I} summarizes the optimal values for the parameters of these two particular forms of $Q$ when determined from the dataset reported in \cite{Shang-2013a} using MATLAB multistart algorithm \cite{MATLAB} and each of the three figures of merit. In addition, the $90\%$ confidence bound of each optimal parameter and the residual error obtained from each optimization are reported in the table. It bears emphasis that the optimization is performed under the assumption that a flat faced tablet and doubly curved tablet have the same tensile strength if they have the same relative density. 
 
\begin{table}[htb]
\caption{A comparison of existing models according to their optimal coefficients, $90\%$ confidence intervals and residual errors obtained from each figure of merit, for all flat and doubly convex tablets. ($\underline{\circ}$ indicates that $\circ$ is an assumption.)}
    \centering
    \scriptsize\setlength{\tabcolsep}{2pt}{
    \begin{tabular}{c|ccc|c|ccc|ccc}
        \hline
        \hline
         Models & $a$ & $b$ & $c$ & $d$ & $e$ & $f$ & $g$ & \rotatebox[origin=c]{45}{$\sigma$-norm} & \rotatebox[origin=c]{45}{$Q$-norm}&\rotatebox[origin=c]{45}{$Q\sigma$-norm}
        \\
        \hline
        \hline
        Pitt's \cite{Pitt-1988}
         & $0.284$ & $0.0126$ & $0.315$ & $0.001$ & $\underline{1}$ & $\underline{1}$ & $\underline{1}$ & $14.96$ & $0.469$&$2.26$
        \\
        \hline
\hline
        1-parameter \cite{Shang-2013a}
        &$0.14$ & $\underline{0}$ & $\frac{1}{2}-a$ & $b$ & $\underline{1}$ & - & $\underline{1}$ & $-$ & $0.260$ & $-$
        \\
        \hline
\hline
        1-parameter
        &$0.1717$ & $\underline{0}$ & $\frac{1}{2}-a$ & $b$ & $\underline{1}$ & - & $\underline{1}$ & $-$ & $0.236$ & $-$
        \\
         & $\pm 4.9\%$ & & & & & &
        \\
        \hline
		1-parameter 
        &$0.1612$ & $\underline{0}$ & $\frac{1}{2}-a$ & $b$ & $\underline{1}$ & - & $\underline{1}$ & $8.11$ & $-$& $-$
		\\
        & $\pm 4.2\%$ & & & & & &
        \\
        \hline		
1-parameter 
        &$0.1530$ & $\underline{0}$ & $\frac{1}{2}-a$ & $b$ & $\underline{1}$ & - & $\underline{1}$ & $-$ & $-$ &$1.07$
		\\
      & $\pm 4.8\%$ & & & & & &
        \\
        \hline	
\hline	
	4-parameter \cite{Shang-2013a}
        &$0.227$ & $-0.00432$ & $0.117$ & $0.0192$ & $\underline{1}$ & $\underline{1}$ & $\underline{1}$ & $-$ & $0.191$& $-$
        \\
 \hline
	4-parameter
       &$0.2256$ & $-0.0033$ & $0.142$ & $0.0192$ & $\underline{1}$ & $\underline{1}$ & $\underline{1}$ & $-$ &$0.164$ &$-$
      \\
       & $\pm 10\%$ &  $\pm 60.4\%$ & $\pm 23.7\%$ &$\pm 36.5\%$  & & &
        \\
        \hline
	4-parameter
       &$0.1224$ & $0.0079$ & $0.3267$ & $-0.0055$ & $\underline{1}$ & $\underline{1}$ & $\underline{1}$ & $6.22$ & $-$& $-$
         \\
       & $\pm 22.9\%$ &  $\pm 46.2\%$ & $\pm 15.1\%$ &$\pm 170.2\%$  & & &
        \\
        \hline
4-parameter
         &$0.1625$ & $0.0027$  & $0.2595$ & $0.0029$ & $\underline{1}$ &$\underline{1}$& $\underline{1}$ & $-$ & $-$ &$0.772$
		\\ 
        & $\pm 14.7\%$ &  $\pm 94.6\%$ & $\pm 15\%$ &$\pm 294.2\%$  & & &
        \\
        \hline
\hline
    \end{tabular}
}
    \label{Table-MaterialProperties-I}
\end{table}

The minimization of the $Q$-norm has been used by previous authors and therefore the corresponding fitted parameters are close to those reported in \cite{Shang-2013a}. It is interesting to note that our optimal values for the parameters result in a smaller residual error than that obtained with previously reported values. This may be attributed to the good performance of the multistart algorithm employed or to rounding errors in the values of $\sigma_t(\rho_R)$ retrieved from \cite{Shang-2013a}. In the case of the 4-parameter model, the improvement over Pitt's equation is evident. These results also reveal that the 1-parameter model leads to a well-defined stable optimization problem (i.e., error bounds are small and solutions are less sensitive to the figure of merit) and that the 4-parameter model leads to a better physical description of the tensile strength (i.e., the residual errors are systematically smaller than those obtained with the 1-parameter model).

It is important to note that the fidelity and robustness of these optimal tensile strength relationships can be further improved by: (i) considering a more general, mechanistically informed expression for $Q$, (ii) extending the size and variety of the experimental dataset, (iii) restricting attention to those tablets which failed under pure tensile stress (see, e.g.,   \cite{Shang-2013a,Shang-2013b,Podczeck-2013} for other failure mechanisms). These three aspects are examined next in turn.

\subsection{General model}

The correct limit of $Q$ for flat tablets, i.e., $Q \rightarrow t/2D$ as $W \rightarrow t$, can be imposed analytically by writing Eq.~\eqref{eq:general} as follows
\begin{equation}
	\sigma_t = \frac{F}{\pi D^2 \left[a\left(\frac{W}{D}\right)^e+ b\left(\frac{t}{W}\right)^f+\frac{1}{2}\left(\frac{t}{D}\right)-a\left(\frac{t}{D}\right)^e-b\right]}
\label{Eq:general}
\end{equation}
where $\{a, b, e, f \}$ are the fitting parameters. We note that \eqref{Eq:general}, referred to as the \emph{general model}, not only exhibits the correct limit for flat geometries and captures the leading order behavior of $Q$ but it also reduces the dimension of the search space from 7 to 4. Coincidentally, the general model and the 4-parameter model (i.e., any re-calibration of Pitt's equation) have a search space of dimension 4. Therefore, the general model requires a computation effort for the optimization of its parameters similar to that of previous models but it allows for a better physical description of the tensile strength. Moreover, by imposing $e=f=1$ on \eqref{Eq:general}---or the correct limit on \eqref{Eq:4-parameter}---a new \emph{2-parameter model} is recovered
\begin{equation}
\sigma_t = \frac{F}{\pi D^2 \left[a\left(\frac{t}{D}\right)+ {b}\left(\frac{t}{W}\right)+ \left(\frac{1}{2}-a\right)\frac{W}{D}-b\right]}
\label{Eq:2-parameter}
\end{equation}
where $\{a, b\}$ are the fitting parameters.

In order to assess the behavior of the proposed models, we restrict attention to those tablets with diameter $D=10.318$~mm reported in  \cite{Shang-2013a} which exhibited crack formation and propagation under pure tensile stress. Specifically, we excluded ball ($D/R = 1.842$), extra deep ($D/R = 1.374$) and some deep tablets ($D/R = 0.988$) having $W/D\approx0.2$ and $t/D\approx0.45$. The optimal values for the parameters of the 1-parameter \eqref{Eq:Shang}, 2-parameter \eqref{Eq:2-parameter}, 4-parameter \eqref{Eq:4-parameter} and general \eqref{Eq:general} models are reported in Table~\ref{Table-MaterialProperties-II}. Confidence bounds and residual errors are also reported in the table.

\begin{table}[htb]
\caption{A comparison between all models according to their optimal coefficients, $90\%$ confidence intervals and residual errors obtained from each figure of merit, for only those flat and doubly convex tablets that failed under pure tensile stress. ($\underline{\circ}$ indicates that $\circ$ is an assumption.)}
    \centering
    \scriptsize\setlength{\tabcolsep}{2pt}{
    \begin{tabular}{c|ccc|c|ccc|ccc}
        \hline
        \hline
         Models & $a$ & $b$ & $c$ & $d$ & $e$ & $f$ & $g$ & \rotatebox[origin=c]{45}{$\sigma$-norm} & \rotatebox[origin=c]{45}{$Q$-norm}&\rotatebox[origin=c]{45}{$Q\sigma$-norm}
        \\
        \hline
  Pitt's \cite{Pitt-1988}
         & $0.284$ & $0.0126$ & $0.315$ & $0.001$ & $\underline{1}$ & $\underline{1}$ & $\underline{1}$ & $13.3$ & $0.43$&$2.13$
        \\
        \hline
\hline
        1-parameter
        &$0.1316$ & $\underline{0}$ & $\frac{1}{2}-a$ & $b$ & $\underline{1}$ & - & $\underline{1}$ & $-$ & $0.196$ & $-$
        \\
         & $\pm 10.1\%$ & & & & & &
        \\
        \hline
		1-parameter 
        &$0.1236$ & $\underline{0}$ & $\frac{1}{2}-a$ & $b$ & $\underline{1}$ & - & $\underline{1}$ & $6.43$ & $-$& $-$
		\\
        & $\pm 7.5\%$ & & & & & &
        \\
        \hline		
1-parameter 
        &$0.1016$ & $\underline{0}$ & $\frac{1}{2}-a$ & $b$ & $\underline{1}$ & - & $\underline{1}$ & $-$ & $-$ &$0.739$
		\\
      & $\pm 9.5\%$ & & & & & &
        \\
        \hline	
\hline
   2-parameter
       &$-0.0202$ & $0.0285$ & $\frac{1}{2}-a$ & $-b$ & $\underline{1}$ & $\underline{1}$ & $\underline{1}$ & $-$ & $0.16$& $-$
       \\
        & $\pm 164.7\%$ &  $\pm 20.7\%$ & & & & &
        \\
        \hline
        2-parameter
        & $-0.0538$& $0.0278$ & $\frac{1}{2}-a$ & $-b$ & $\underline{1}$ & $\underline{1}$ & $\underline{1}$ & $4.71$ & $-$& $-$
          \\
         & $\pm 51.3\%$ &  $\pm 16\%$ & & & & &
        \\
        \hline
 2-parameter
        & $-0.0550$& $0.0278$ & $\frac{1}{2}-a$ & $-b$ & $\underline{1}$ & $\underline{1}$ & $\underline{1}$ & $-$ & $-$& $0.568$
          \\
        & $\pm 50.1\%$ &  $\pm 16.9\%$ & & & & &
        \\
        \hline
\hline
	4-parameter
        &$0.1406$ & $0.0077$ & $0.2626$ & $-0.0037$ & $\underline{1}$ & $\underline{1}$ & $\underline{1}$ & $-$ & $0.139$ &$-$
      \\
       & $\pm 35\%$ &  $\pm 107.7\%$ & $\pm 28.8\%$ &$\pm 430.5\%$  & & &
        \\
        \hline
	4-parameter
       &$-0.0382$ & $0.0296$ & $0.5615$ & $-0.0390$ & $\underline{1}$ & $\underline{1}$ & $\underline{1}$ & $4.5$ & $-$& $-$
         \\
       & $\pm 114.8\%$ &  $\pm 24.9\%$ & $\pm 13.3\%$ &$\pm 37.2\%$  & & &
        \\
        \hline
4-parameter
         &$-0.0132$ & $0.0262$ & $0.5193$ & $-0.0332$ & $\underline{1}$ & $\underline{1}$ & $\underline{1}$ & $-$ & $-$ &$0.539$
      \\
        & $\pm 328.5\%$ &  $\pm 29.5\%$ & $\pm 14\%$ &$\pm 46.5\%$  & & &
        \\
        \hline
\hline
	general
        &$0.3231$ & $0.0240$ & $a$ & $\frac{1}{2}\frac{t}{D}-b$ & $1.6963$ & $-263.5$ & $e$ & $-$ & $0.146$ &$-$
        \\
       & $\pm 10.9\%$ &  $\pm 36\%$ &  &  & $\pm 12\%$&$\pm \infty\%$ &
        \\
        \hline
        general
       & $0.0484$ & $0.8688$ & $a$ & $\frac{1}{2}\frac{t}{D}-b$ & $-0.8553$ & $-0.3147$ & $e$ & $3.85$ & $-$ & $-$
       \\
        & $\pm 267.6\%$ &  $\pm 83\%$ &  &  & $\pm 82.7\%$&$\pm 43.4\%$ &
        \\
        \hline
        general
        & $0.7289$ & $2.3185$ & $a$ & $\frac{1}{2}\frac{t}{D}-b$ & $-0.3151$ & $-0.2156$ & $e$ &$-$ & $-$&$0.463$
        \\
  & $\pm 689.1\%$ &  $\pm 409\%$ &  &  & $\pm 283.8\%$&$\pm 209.9\%$ &
        \\
        \hline
\hline
    \end{tabular}
}
    \label{Table-MaterialProperties-II}
\end{table}

The residual errors for 1-parameter and 4-parameter models are noticeably reduced in comparison to Table~\ref{Table-MaterialProperties-I}, confirming that those tablets that did not fail under tensile stress are outliers of the optimal tensile strength relationships proposed. Furthermore, it is evident from the table that the 2-parameter outperforms the 1-parameter model, suggesting that $t/W$ is required in the expression for $Q$---though perhaps to a power $f$ different from 1. Finally, the 4-parameter and the general models have very similar residual errors for all three figures of merit and, in particular, the 4-parameter model exhibits a better performance for the $Q$-norm. 

This last observation provides additional insight into the role of flat faced tablets in the optimization process. The 4-parameter model does not have the correct limiting behavior for flat faced tablets. However, only 5.7\% are flat faced tablets in the dataset and thus their contribution to the overall residual error is negligible.  In other words, the optimization process reduces the error for doubly convex tablets in detriment to the predictability of the model for shallow/flat tablets. Specifically, the 4-parameter model exhibits, in average, a 65$\%$ larger error for flat faced tablets and a 5$\%$ smaller error for curved tablets than the general model. The inclusion of more flat tablets in the dataset may, however, have the opposite effect. We further study this issue in the next subsection.

\subsection{Role of flat faced tablets in the optimization process}
\label{Section:FlatTablets}

\begin{table}[htb]
    \centering
\caption{Recalibration of optimal coefficients, $90\%$ confidence interval and residual errors for flat and doubly convex tablets that failed under pure tensile stress, using a larger group of flat faced tablets in the optimization process. ($\underline{\circ}$ indicates that $\circ$ is an assumption.)}
  \scriptsize\setlength{\tabcolsep}{2pt}{
    \begin{tabular}{c|ccc|c|ccc|ccc}
      \hline
\hline
         Models & $a$ & $b$ & $c$ & $d$ & $e$ & $f$ & $g$ & \rotatebox[origin=c]{45}{$\sigma$-norm} & \rotatebox[origin=c]{45}{$Q$-norm}&\rotatebox[origin=c]{45}{$Q\sigma$-norm}
    \\
\hline
\hline
        1-parameter 
       &$0.1145$ & $\underline{0}$ & $\frac{1}{2}-a$ & $b$ & $\underline{1}$ & - & $\underline{1}$ & $-$ & $0.229$& $-$
        \\
        &$\pm 12.4\%$ & &  & & & & & &
        \\
        \hline
	 1-parameter 
         &$0.0949$ & $\underline{0}$ & $\frac{1}{2}-a$ & $b$ & $\underline{1}$ & - & $\underline{1}$ & $8.95$ & $-$
	      \\
        &$\pm 11\%$ & &  & & & & & &
        \\
        \hline
        1-parameter 
        &$0.0654$ & $\underline{0}$ & $\frac{1}{2}-a$ & $b$ & $\underline{1}$ & - & $\underline{1}$ & $-$ & $-$ &$1.025$
	\\
 &$\pm 17.7\%$ & &  & & & & & &
        \\
        \hline
\hline
        2-parameter
          &$-0.0380$ & $0.0287$ & $\frac{1}{2}-a$ & $-b$ & $\underline{1}$ & $\underline{1}$ & $\underline{1}$ & $-$ & $0.200$ &$-$
        \\
        & $\pm 99\%$ &  $\pm 23.3\%$ & & & & &
        \\
        \hline
        2-parameter
         &$-0.1117$ & $0.0321$ & $\frac{1}{2}-a$ & $-b$ & $\underline{1}$ & $\underline{1}$ & $\underline{1}$ & $6.90$ & $-$ &$-$
        \\
 & $\pm 29.4\%$ &  $\pm 16.6\%$ & & & & &
        \\
        \hline
   2-parameter
     &$-0.1205$ & $0.0330$ & $\frac{1}{2}-a$ & $-b$ & $\underline{1}$ & $\underline{1}$ & $\underline{1}$ &$-$ & $-$ & $0.838$
        \\
 & $\pm 29.1\%$ &  $\pm 18.2\%$ & & & & &
        \\
        \hline
\hline
	4-parameter
        &$0.0442$ & $0.0198$ & $0.4183$ & $-0.0185$ & $\underline{1}$ & $\underline{1}$ & $\underline{1}$ & $-$ & $0.187$ &$-$
        \\
 & $\pm 109\%$ &  $\pm 46.6\%$ &$\pm 18.1\%$ &$\pm 99.5\%$ & & &
        \\
        \hline
	4-parameter
        &$-0.1187$ & $0.0415$ & $0.6975$ & $-0.0655$ & $\underline{1}$ & $\underline{1}$ & $\underline{1}$ & $6.26$ & $-$& $-$
        \\
& $\pm 31.2\%$ &  $\pm 15.7\%$ &$\pm 7.9\%$ &$\pm 18.8\%$ & & &
        \\
        \hline
	4-parameter
        &$-0.1316$ & $0.0444$ & $0.7211$ & $-0.0720$ & $\underline{1}$ & $\underline{1}$ & $\underline{1}$ & $-$ & $-$ & $0.776$
        \\
& $\pm 29.2\%$ &  $\pm 16.6\%$ &$\pm 8\%$ &$\pm 20.8\%$ & & &
        \\
        \hline
\hline
        general 
       & $3.0992$ & $3.9923$ & $a$ & $\frac{1}{2}\frac{t}{D}-b$ & $-0.1640$ & $-0.2006$ & $e$ &$-$  & $0.178$ & $-$
        \\   
  & $\pm 1313\%$ &  $\pm 972.5\%$ &  &  & $\pm 593.1\%$&$\pm 508.3$ &
        \\
        \hline
       general 
       & $0.0342$ & $0.6553$ & $a$ & $\frac{1}{2}\frac{t}{D}-b$ & $-0.9648$ & $-0.436$ & $e$ & $5.54$ & $-$ & $-$
        \\   
  & $\pm 278.9\%$ &  $\pm 76.4\%$ &  &  & $\pm 82\%$&$\pm 41.1$ &
        \\
        \hline
	 general
       & $0.1344$ & $0.8150$ & $a$ & $\frac{1}{2}\frac{t}{D}-b$ & $-0.6065$ & $-0.4411$ & $e$ &  $-$ & $-$ &$0.687$
        \\   
  & $\pm 517.8\%$ &  $\pm 199.8\%$ &  &  & $\pm 182.4\%$&$\pm 186.3$ &
        \\
        \hline
\hline
    \end{tabular}
    }
    \label{Table-MaterialProperties-new}
\end{table}

We extended the experimental dataset in \cite{Shang-2013a} with a new series of tests on flat faced tablets. Specifically, 28 flat tablets of pure microcrystalline cellulose (Avicel Ph102), with true density 1540 kg/m$^3$, were manufactured using a 10~mm flat face B-tooling in a linear compaction emulator (Presster, Metropolitan Computing Corp., NJ). The tablets were diametrically compressed using an Instron testing machine at a loading rate of 10 mm/min (see Table~{\ref{tab:flat tablets} in the Appendix for numerical values). All tablets exhibited failure under pure tensile stress. The new fitting parameters to the exponential function \eqref{eq:RD} are $A=20.65$~kPa and $B=6.787$, as depicted in Fig.~\ref{Fig: stress-RD}.

Table~\ref{Table-MaterialProperties-new} shows the optimal values for the parameters of the 1-parameter \eqref{Eq:Shang}, 2-parameter \eqref{Eq:2-parameter}, 4-parameter \eqref{Eq:4-parameter} and general \eqref{Eq:general} models when calibrated with the extended experimental campaign. Flat faced tablets now represent 21.3\% of the total number of tablets (cf. 5.7\% in the previous Section). In contrast to results in Table~\ref{Table-MaterialProperties-II}, the general model exhibits smaller residual errors than those of the 4-parameter model for all the figures of merit. This result confirms that, by including more flat faced tablets in the dataset, the limiting behavior of the 4-parameter model is improved only in detriment of its overall behavior. Specifically, the 4-parameter model now exhibits, in average, a 1$\%$ smaller error for flat faced tablets and a 17$\%$ larger error for curved tablets than the general model. The general model, however, automatically exhibits the correct limit, rendering unnecessary the need of a large number of experiments for flat geometries. The cost- and time-effectiveness of using the general model is evident.

It bears emphasis that experimental errors and uncertainty in the functionality of the geometric function $Q$ render the problem ill-posed (i.e., the solution is not unique, sensitive to errors, and dependent on the norm which is minimized). Experimental errors cannot be eliminated but one can minimize the figure of merit that provides more stability to the optimization process. According to our case study, this is the case of the $\sigma\mathrm{-norm}$ and thus the optimization problem reduces to
\begin{linenomath*}
\begin{eqnarray*}
	\min_{a, b, e, f}
		\left[ 
			\sum_{i \in \mathcal{P}}  
			\left( \sigma_t(\rho_i) - \frac{F_i}{\pi~D_i^2~\left[a\left(\frac{W_i}{D_i}\right)^e+ b\left(\frac{t_i/D_i}{W_i/D_i}\right)^f+\frac{1}{2}\left(\frac{t_i}{D_i}\right)-a\left(\frac{t_i}{D_i}\right)^e-b\right]} \right)^2 
		\right]^{1/2} 
\end{eqnarray*}
\end{linenomath*}
where $\mathcal{P}$ is a set of experimental points and $\sigma_t(\rho_i)$ is obtained from a small number of flat faced tablets.

\subsection{Mechanistic interpretation}

A major source of uncertainty is the fact that the functionality of $Q$ is unknown in general. However, further insight can be gained by recasting the problem in terms of an effective cross-sectional surface area, $\bar{A}$, associated with strength, that is
\begin{equation}
	\sigma_t = \frac{2 F}{\pi \bar{A}}
\end{equation}
where $\bar{A}=t D$ for flat-faced tablets (cf. Eq.~\eqref{Eq:Hertz}). For doubly convex tablets, we parametrize $\bar{A}$ by an effective thickness $\bar{t}$ (see Fig.~\ref{Fig:Tablet-failure}) as follows
\begin{equation}
 \bar{A} = D^2 \left[2\left(\frac{\bar{t}}{D}-\frac{W}{D}\right)\left[\frac{1}{3}+\frac{1}{15} \left( \frac{\bar{t}}{D}-\frac{W}{D}\right)^2\right]+\frac{W}{D}\right]
             +
             \mathcal{O}\left(\frac{\left(\bar{t}-W\right)^5}{D^3} \right)             
\label{eq:cross-section}
\end{equation}
This parametrization is made only in the interest of simplicity. However, a geometric interpretation of the above equation suggests that $\bar{t}/D$ may be a function of $D/R$, which we postulate to be 
\begin{linenomath*}
\begin{eqnarray*}
\frac{\bar{t}}{D} = \frac{W}{D} + \alpha \left(\frac{D}{R}\right)^\beta \left(\frac{t}{D}-\frac{W}{D}\right)
\end{eqnarray*}
\end{linenomath*}
where $\alpha > 0$ and $\beta \ge 0$ are fitting parameters. As a result, a new relationship between geometric parameters, breaking force and tensile strength is obtained, i.e., 
\begin{equation}
	\sigma_t = \frac{F}{\pi D^2  \left[2\alpha\left(\frac{D}{R}\right)^\beta\left(\frac{t}{D}-\frac{W}{D}\right)\left[\frac{1}{3}+\frac{\alpha^2}{15}\left(\frac{D}{R}\right)^{2\beta}\left(\frac{t}{D}-\frac{W}{D}\right)^2\right]+\frac{W}{D}\right]}
\label{Eq:mechanistic}
\end{equation}
which is referred to as the \emph{mechanistic model}. It is interesting to note that the mechanistic model has a substantially different functionality compared to the one of the general model \eqref{Eq:general}.In the case of $\bar{t} = (t+W)/2$, however, the 1-parameter model is approximately recovered with $a=1/6$, which is very close to the optimal value obtained in our case study (see Table~\ref{Table-MaterialProperties-I}). Thus, a more clear connection between the mechanistic model and the general model is desirable, if beyond the scope of this work. 

The optimal values for $\alpha$ and $\beta$, determined from the extended dataset (Section~\ref{Section:FlatTablets}) using each of the three figures of merit, are presented in Table~\ref{Table-cross-section}. The residual errors are comparable to those of the general model and the stability is remarkable, as shown by the tight confidence bounds. The optimal values are clearly insensitive to the figure of merit. In addition, the search space of the mechanistic model is of dimension 2 whereas the one of the general model is of dimension 4. Thus, the mechanistic model is preferable both in terms of efficacy and efficiency.

\begin{table}[htb]
\caption{Mechanistic model optimal coefficients, $90\%$ confidence bound and residual errors from each figure of merit, for those tablets in extended dataset that failed under pure tensile stress.}
    \centering
    \scriptsize{
    \begin{tabular}{c|cc|ccc}
        \hline
        \hline
         Model & $\alpha$ & $\beta$ & \rotatebox[origin=c]{45}{$\sigma$-norm} & \rotatebox[origin=c]{45}{$Q$-norm}&\rotatebox[origin=c]{45}{$Q\sigma$-norm}
        \\
	\hline
\hline
    mechanistic
       &$0.5817$ & $3.9736$ & $-$ & $0.172$ & $-$ 
        \\
        & $\pm 8.6\%$& $\pm 25.5\%$
        \\
        \hline
        mechanistic
       &$0.5562$ & $3.9877$ & $5.97$& $-$ & $-$ 
        \\
        & $\pm 9.4\%$& $\pm 15.4\%$
        \\
        \hline
   mechanistic
      &$0.5377$ & $5.0030$ & $-$ & $-$ &$0.699$
        \\
        & $\pm 9.4\%$& $\pm 18.3\%$
        \\
     \hline
\hline
  \end{tabular}
    }
    \label{Table-cross-section}
\end{table}

\subsection{Recommendations and best practices}

The proposed approach provides a general framework for determining optimal relationships for tensile strength of doubly convex tablets under diametrical compression. Under this framework, other expressions for the nonlinear function $Q$ can be explored and the assumption that $Q$ only depends on geometric parameters can even be relaxed. Similarly, figures of merit other than those studied here (i.e., $\sigma$-norm, $Q$-norm and $Q\sigma$-norm) can be examined. It bears emphasis that, regardless of the choice of $Q$ and the optimization procedure, the performance of a new model can be assessed following the same procedure presented here. We illustrate the guidelines for this procedure by investigating the performance of the five models described in this work:
\begin{itemize}
\item Number of parameters: If the number of parameters (i.e., dimension of the search space) is small, then the optimization process is less computationally expensive. The third column in Table~\ref{Table:models-summary} indicates that the 1-parameter, 2-parameter and mechanistic models are preferable.
\item Limiting behavior for flat faced tablets: If the correct limiting behavior for flat geometries is automatically provided by the model, then $\sigma_t(\rho_i)$ can be obtained from a small number of flat faced tablets---otherwise, a larger number of flat tablets is required in the optimization of the model's parameters. The fourth column in Table~\ref{Table:models-summary} shows that the 4-parameter model is the only model that doesn't provide the correct limiting behavior.
\item Stability of the optimization process: If the optimization process is stable then the confidence bounds of the optimal value for each parameter are tighter. Narrow confidence intervals tell that the estimated value is relatively stable. The confidence bounds reported in Table \ref{Table-MaterialProperties-new} allow the comparison of stability between the models. The fifth column in Table~\ref{Table:models-summary} reveals that the 1-parameter and mechanistic models are the most stable. 
\item Predictability of the model: If the residual error obtained from the optimization is small, then the predictability of the model is high. In addition, the distribution of the errors that contribute to the overall residual can be used to further discern between models. The sixth column in Table~\ref{Table:models-summary} indicates that the general and mechanistic models are the most predictive. Furthermore,  Figure~\ref{Fig:all-data} shows that the mechanistic model has the largest number of small errors.
\end{itemize}

It is evident from this analysis that the general and mechanistic models are preferable over other optimal relationships for tensile strength. Moreover, the mechanistic model is the most stable with only 2 parameters to be optimized, namely $\alpha$ and $\beta$ in Eq.~\eqref{Eq:mechanistic}. 

\begin{table}[htb]
\caption{A systematic comparison of the proposed and existing models based on four different performance criteria.}
    \centering
    \scriptsize{
    \begin{tabular}{c|c|cc|cc}
        \hline
        \hline
        Model & Equation & Number of  & Limiting  & Stability & Predictability
        \\
                   &   & Parameters  & Behavior  & (1:best--5:worst) & (1:best--5:worst)
        \\
	\hline
	\hline
	1-parameter
	& \eqref{Eq:Shang} & $1$ & $\checkmark$ & 1({$\sigma$-norm}) & 5 
        \\
	2-parameter
	& \eqref{Eq:2-parameter} & $2$ & $\checkmark$ & 3 ({$\sigma$, Q$\sigma$-norm})  & 4 
        \\
	4-parameter
	& \eqref{Eq:4-parameter} & $4$ & $\times$ & 4 ({$\sigma$, Q$\sigma$-norm})  & 3
        \\
	general
	& \eqref{Eq:general} & 4 & $\checkmark$ & 5 ({$\sigma$-norm}) &  1
	\\
	mechanistic
	& \eqref{Eq:mechanistic} & 2 & $\checkmark$ & 2 ($\sigma$-norm) & 2
	\\
\hline
  \end{tabular}
    }
    \label{Table:models-summary}
\end{table}

\begin{figure}[htb]
\centering
    \begin{tabular}{cc}
        \includegraphics[scale=0.41]{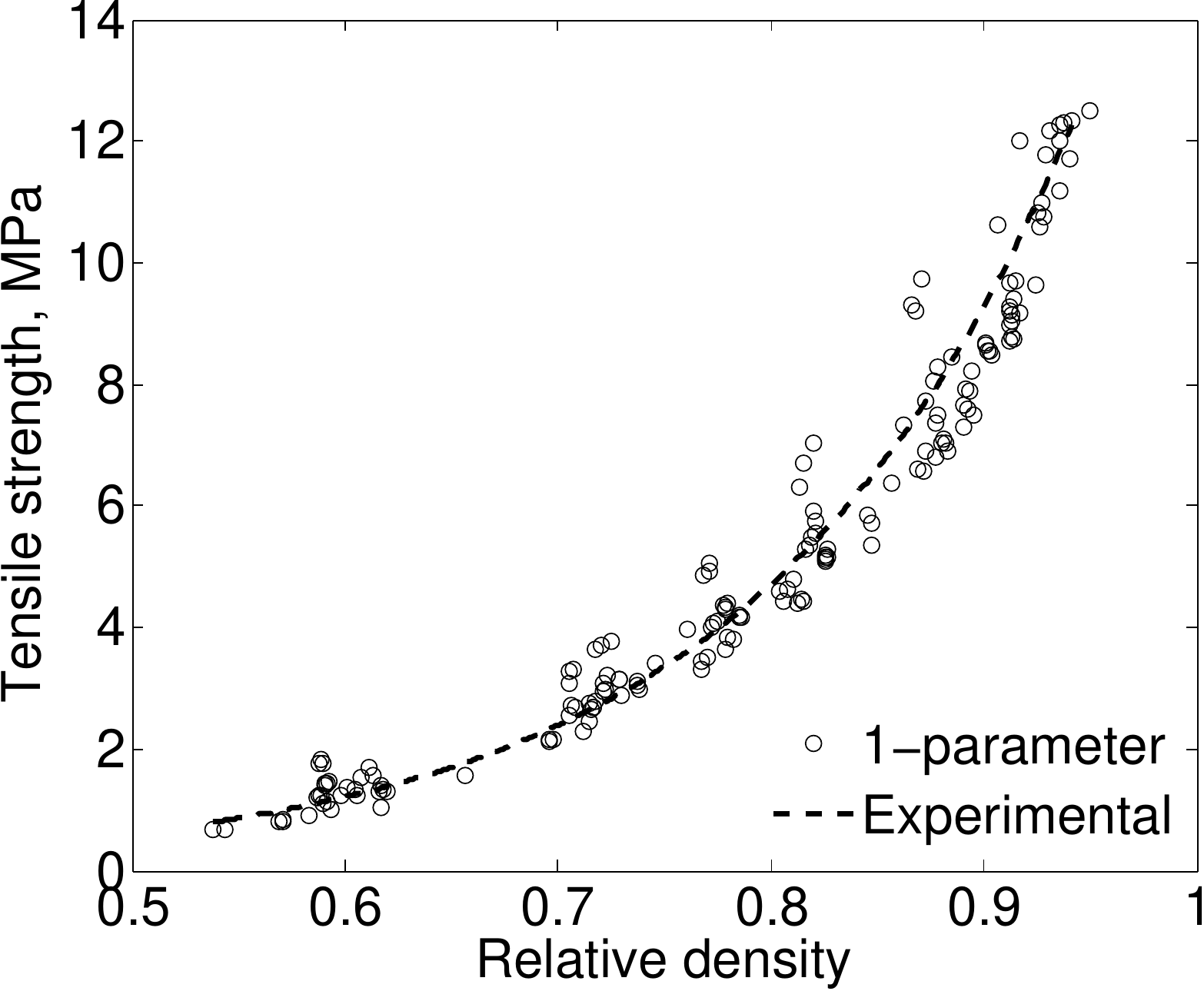}
                \Put(-163,215){\includegraphics[scale=0.2]{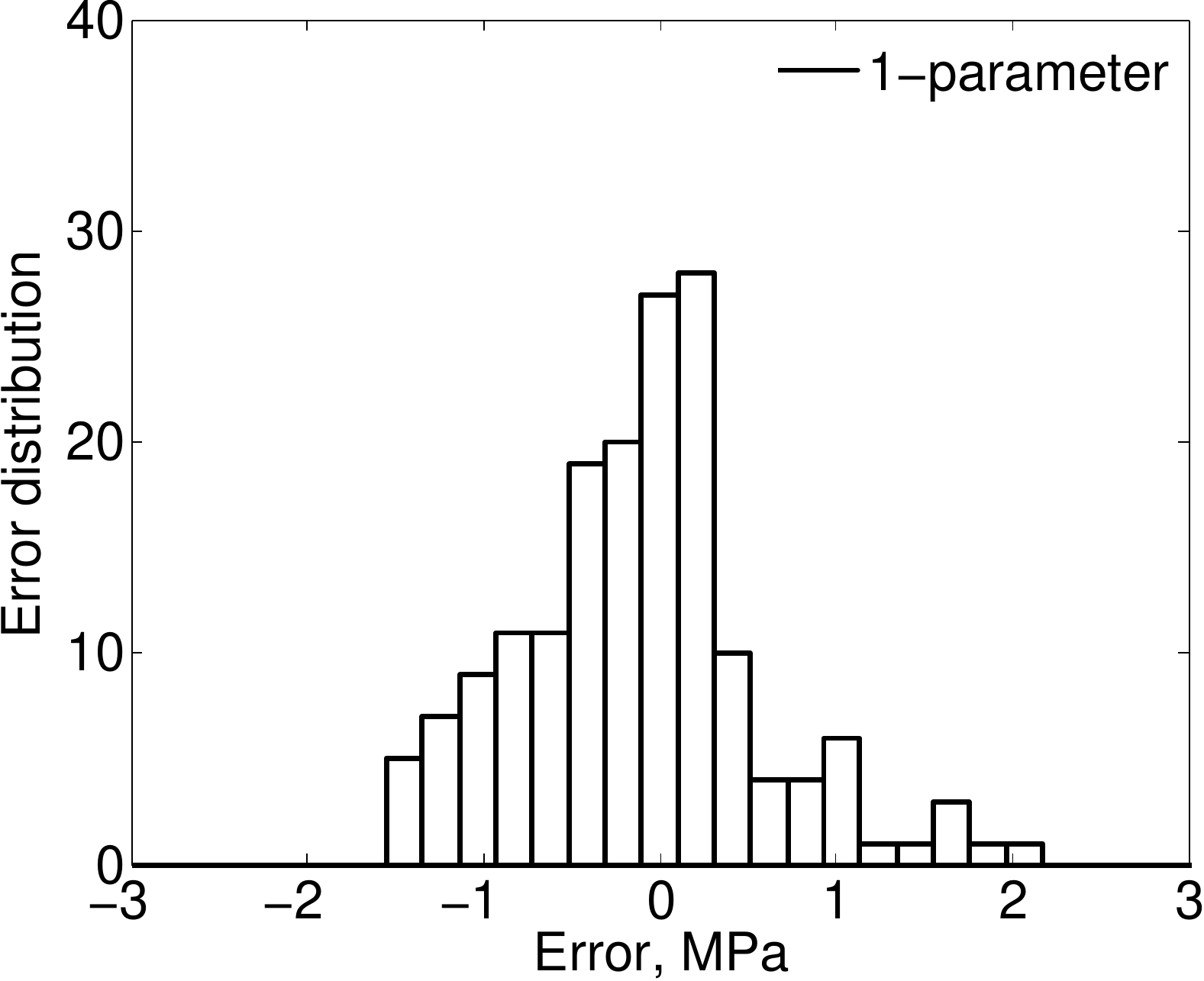}}
        &
        \includegraphics[scale=0.41]{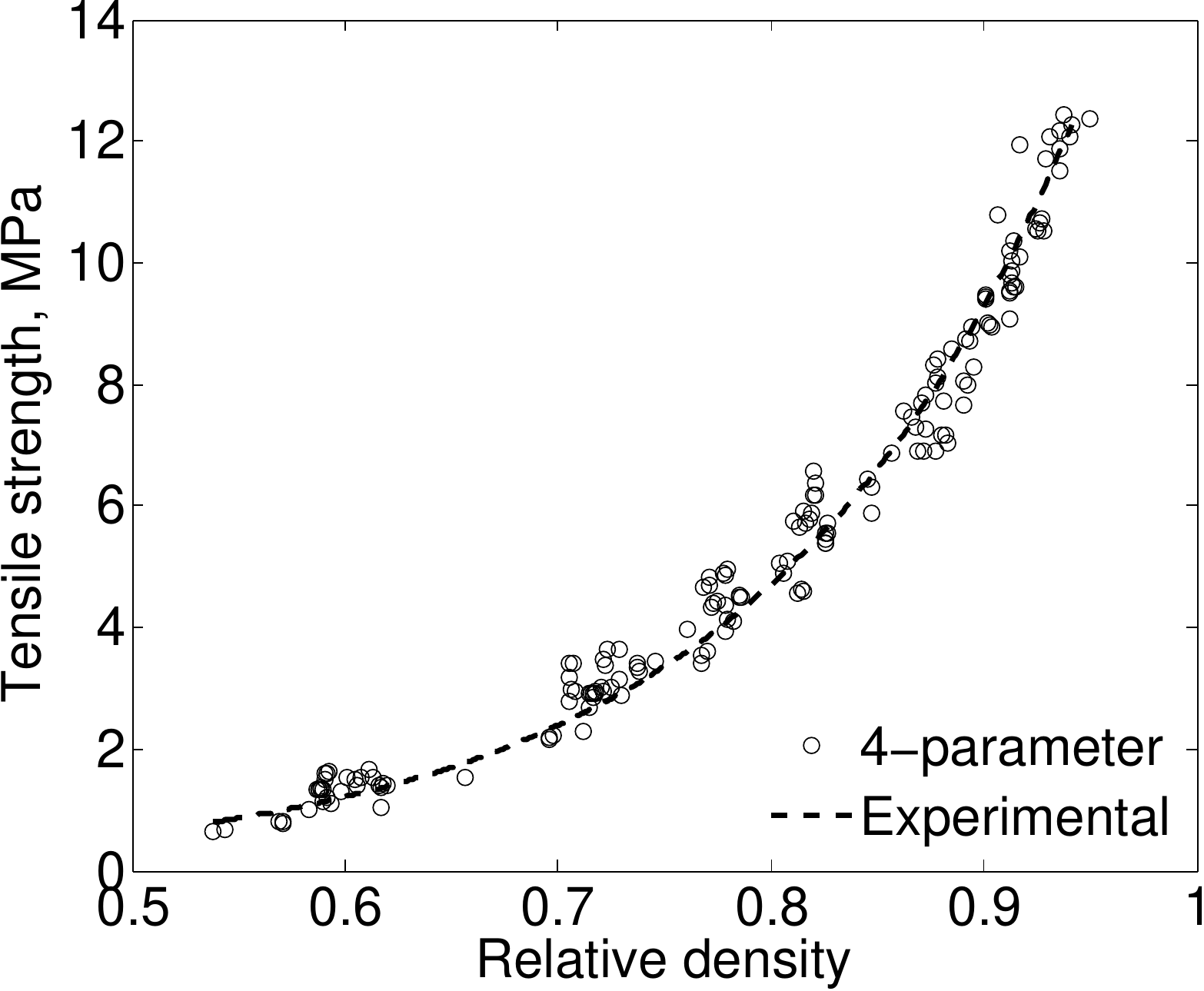}
		\Put(-163,215){\includegraphics[scale=0.2]{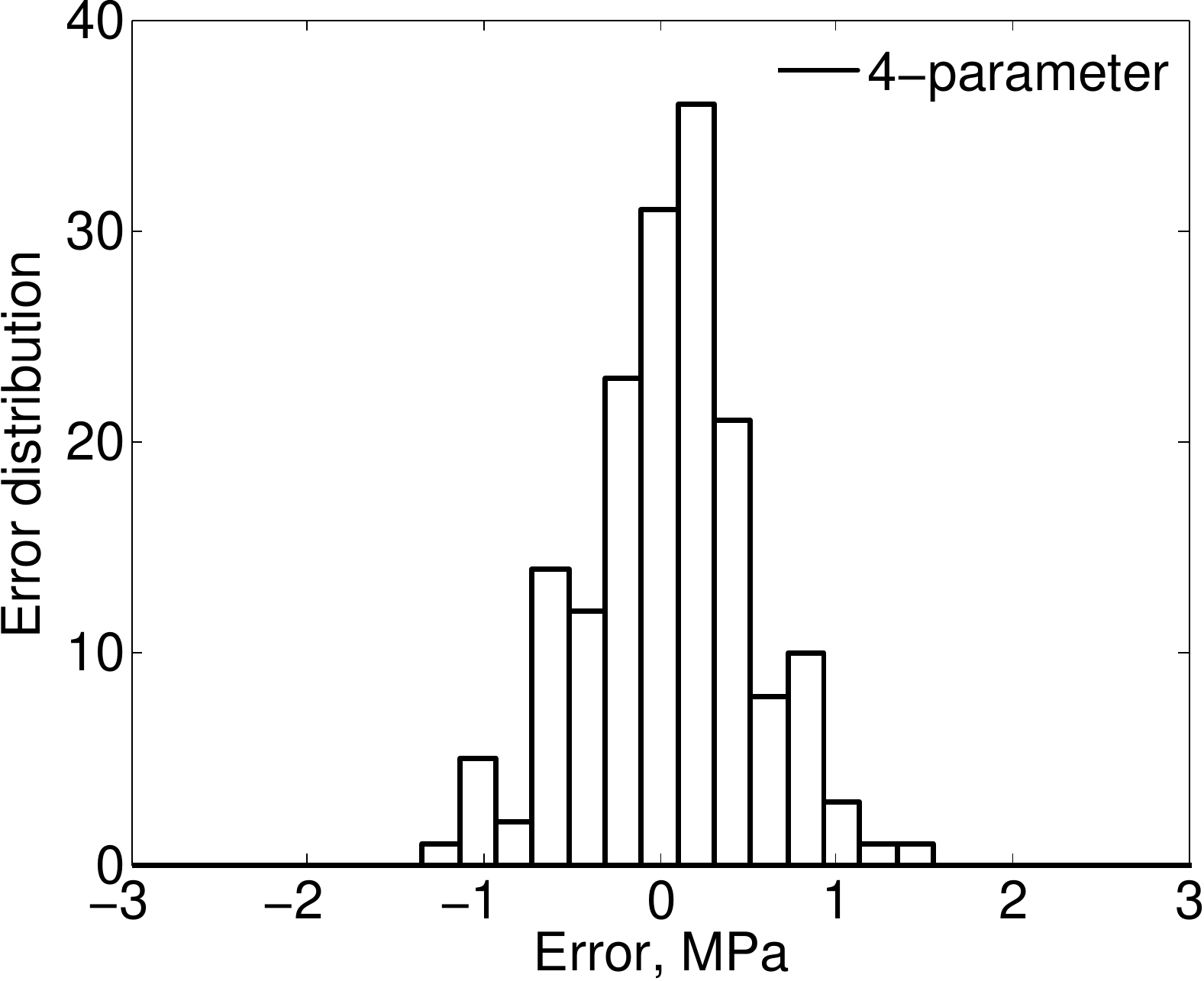}}
        \\
        \includegraphics[scale=0.41]{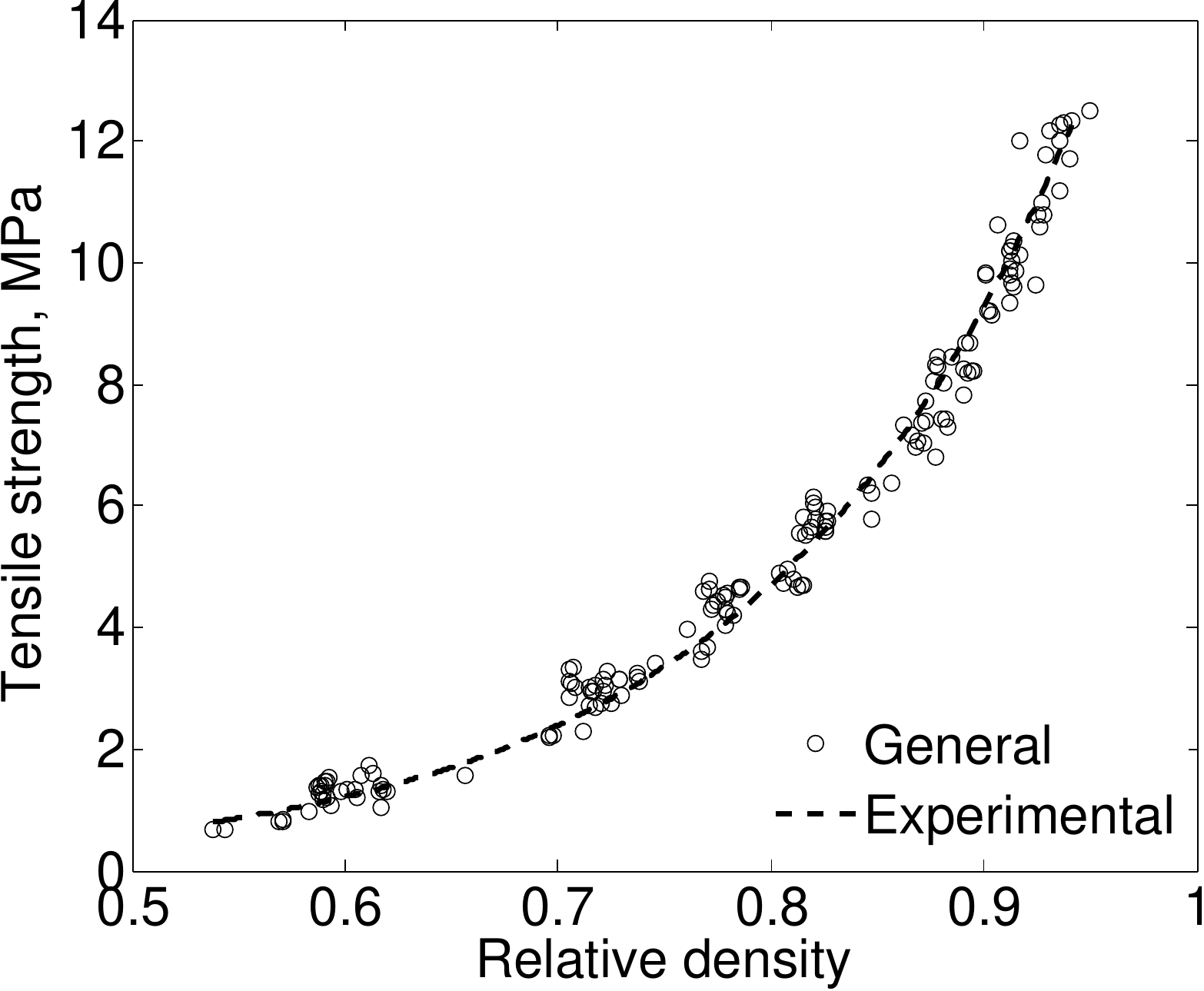}
		\Put(-163,215){\includegraphics[scale=0.2]{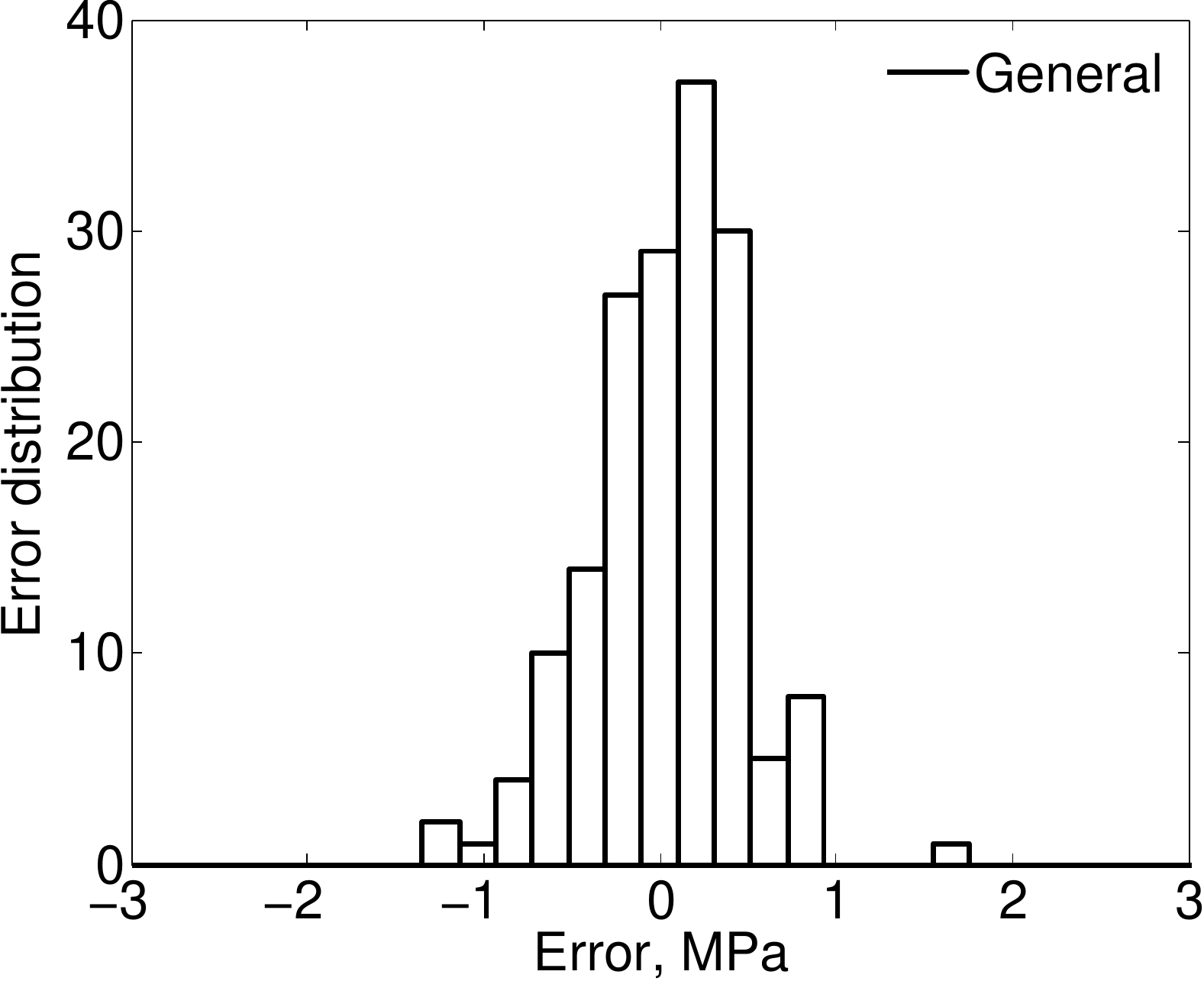}}
        &       
        \includegraphics[scale=0.41]{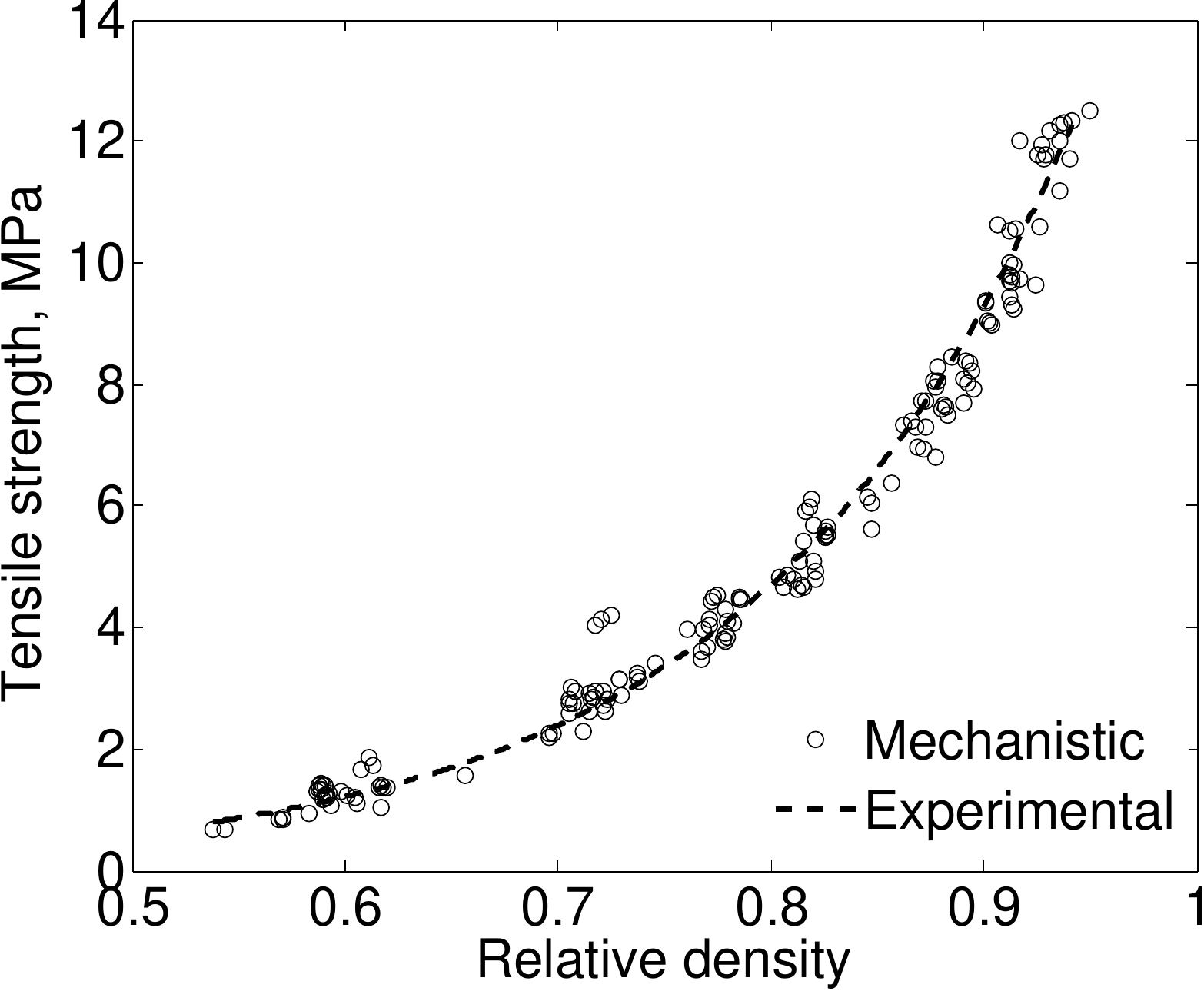}
		\Put(-163,215){\includegraphics[scale=0.2]{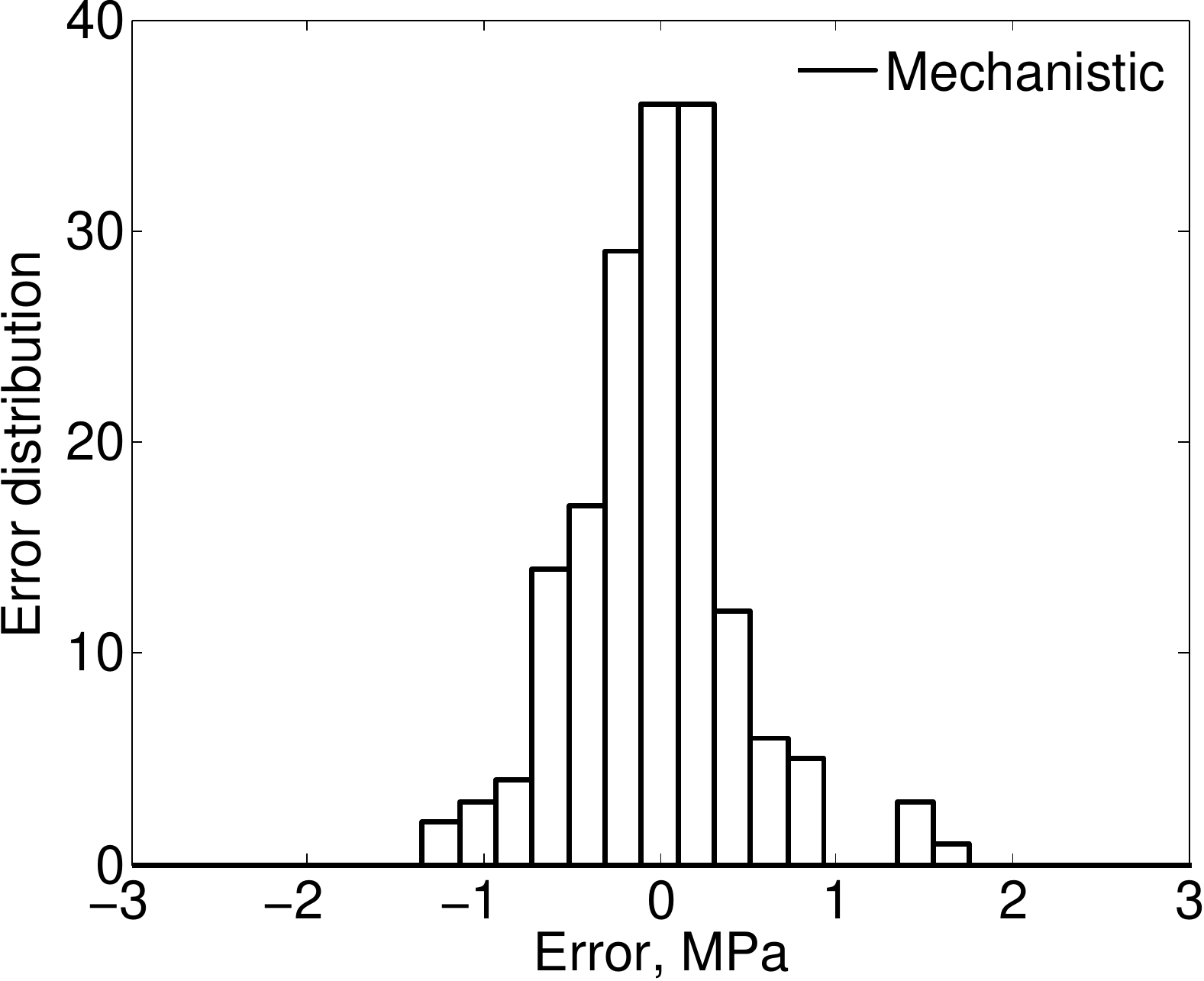}}
        \end{tabular}   
\caption{The relationship between tensile strength and relative density of the experimental data using $\sigma\mathrm{-norm}$ along with their corresponding error distribution.}
\label{Fig:all-data}
\end{figure}

\section{Summary and discussion}

We have proposed a general framework for determining optimal relationships for tensile strength of doubly convex tablets under diametrical compression. The approach is based on the observation that tensile strength is directly proportional to the breaking force and inversely proportional to $Q$, a nonlinear function of geometric parameters and materials properties. This generalization reduces to the analytical expression commonly used for flat faced tablets, i.e., Hertz solution, for $Q=t/2D$. Here, we have assumed that $Q$ is solely a function of geometric parameters that, for doubly convex tablets, reduce to $t/D$ and $W/D$, and a combination thereof. Based on such assumptions, the function $Q$ can be determined from experimental results by solving an optimization problem that minimizes a figure of merit of choice. We have postulated that this figure of merit has to be based on the assumption that a flat faced tablet and doubly curved tablet have the same tensile strength if they have the same relative density and are made of the same powder, under equivalent manufacturing conditions. 

In addition, we have presented guidelines for assessing the performance of optimal tensile strength relationships. A desirable model needs to have a small number of optimal parameters and the correct analytical limiting behavior for flat faced tablets (i.e., Hertz solution). It has to be predictive, i.e., the optimization has to result in a narrow and symmetric distribution of errors around zero, for a given figure of merit. This figure of merit in turn has to render the optimization problem stable, i.e., it has to provide optimal parameters with tight confidence bounds. However, it bears emphasis that experimental errors and uncertainty in the functionality of the geometric function $Q$ may render the problem ill-posed. Therefore, these guidelines have to be followed with caution to resolve ambiguities in the results.

We have specifically investigated three different figures of merit, which we referred to as $\sigma$-norm, $Q$-norm and $Q\sigma$-norm, and we have proposed three new optimal tensile strength relationships, which we referred to as \emph{general model}, \emph{2-parameter model} and \emph{mechanistic model}. The general model captures the leading order behavior of $Q$ on the geometric parameters, it has the exact limiting behavior for flat faced tablets, and it has four optimal parameters to be determined. The 2-parameter model simplifies the general model by assuming that $Q$ is linear on the geometric parameters, and thus the number of parameters is reduced to two while the correct limiting behavior is retained. The mechanistic model is based on an effective cross-sectional surface area associated with strength (i.e., in contrast to previous models, it has a well-defined mechanistic interpretation), it exhibits the exact limiting behavior for flat geometries, and it only has two optimal parameters. 

We have assessed the performance of the proposed new models together with two models previously proposed in the literature, i.e., Shang's model (a 1-parameter model introduced in \cite{Shang-2013a}) and a 4-parameter model (based on the model introduced by Pitt \cite{Pitt-1988} which is widely used in the pharmaceutical industry \cite{Pharmacopeia-2011}). This study shows that the general model and the mechanistic model are more predictive than previously proposed tensile strength relationships. Both models automatically exhibit the correct limit for flat geometries, thus only a small number of flat faced tablets has to be tested in order to accurately capture the strength-relative density relationship. Our analysis also indicates that the mechanistic model is the most stable among the predictive models. This is in sharp contrast to the 4-parameter model, i.e., a re-calibration of Pitt's equation, that leads to an unstable optimization problem which, in addition, requires a large number of flat faced tablets in order to remain predictive in the limit of shallow/flat tablets.

These observations suggest that both general and mechanistic models are cost- and time-effective, predictive alternatives to the tensile strength relationship currently used in the pharmaceutical industry. Furthermore, our analysis showcases the benefits of adopting a general framework for developing and evaluating the performance of optimal relationships for tensile strength of doubly convex tablets under diametrical compression.

We close by pointing out some limitations of our analysis and possible avenues for extensions of the general framework.

First, it is clear that the proposed parametric approximations for $Q$ are not the only nonlinear functions of geometric parameters that exhibit the correct limit for flat faced tablets. In addition, tensile strength of flat faced tablets may not be optimally described by an exponential function of the relative density. It is also possible that the function $Q$ has to depend on powder properties or manufacturing variables in some cases of industrial relevance. The systematic investigation of functions $Q$ of the type proposed here, the elucidation of their properties and the determination of the best optimal relationships in each area of application, are worthwhile directions of future research.

Second, our general framework relies in the assumption that, for given powder and manufacturing conditions, a flat faced tablet and doubly curved tablet have the same tensile strength if they have the same relative density. This is indeed a good first order approximation for the case study presented here (i.e., pure microcrystalline cellulose pressed at low compaction speeds). This is probably not true in general, however there is no experimental technique capable to directly address this question (i.e., most techniques perform indirect measurements of effective properties). Thus, particle mechanics simulations capable of describing strength formation and evolution during the compaction process are desirable \cite{Gonzalez-2012,Gonzalez-2014a,Gonzalez-2014b}, if beyond the scope of this paper.

\section*{Conflict of interest}

The authors confirm that there are no conflicts of interest.

\section*{Acknowledgements}

This material is based upon work supported by the National Science Foundation under Grant number IIP-1237873, Industry-Academia Research Partnership for Developing \& Implementing Non-Destructive Characterization and Assessment of Pharmaceutical Oral Dosages in Continuous Manufacturing. The authors also gratefully acknowledge the support received from the NSF ERC grant number EEC-0540855, ERC for Structured Organic Particulate Systems.

\section*{Appendix A.}

\begin{table}[htb]
\begin{center}
\renewcommand\thetable{A.1}
\caption{Tensile strength and relative density of flat faced tablets reported in \cite{Shang-2013a}.}
\vspace{.5pc}
    \scriptsize{
\begin{tabular}{c c}
\hline
\hline
Relative density & Tensile strength, MPa \\
\hline
\hline
0.5910 & 1.388\\
0.7290 & 3.1453\\
0.8103 & 4.7896\\
0.8568 & 6.3768\\
0.8942 & 8.2184\\
0.9239 & 9.6353\\
0.9355 & 11.1944\\
0.9406 & 11.7044\\
\hline
\hline
\label{tab: stress-RD}
\end{tabular}
}
\end{center}
\end{table}

\begin{table}[htb]
\begin{center}
\renewcommand\thetable{A.2}
\caption{Characteristics of flat faced tablets and their calculated tensile strength values.}
\vspace{.5pc}
\scriptsize{
\begin{tabular}{c c c c c c}
\hline
\hline
&Actual weight (g) & Thickness (mm) & Actual diameter (mm) & Break force (N) & Tensile strength (MPa)\\
\hline
\hline
1&0.2969 & 2.79&10.04 &339.39 &7.7133\\
2&0.272 & 2.55&10.03&323.12&8.0427\\
3&0.2718&2.58&10.05&299.14 &7.3446\\
4&0.2706 & 2.84&10.06&193.097&4.3027\\
5&0.268 & 3&10.06&136.74&2.8844\\
6&0.2687& 3.03&10.08&141.41&2.9475\\
7&0.267 & 3.42&10.08&84.2&1.5549\\
8&0.2428 & 3.66&10.1&38.69&0.6663\\
9&0.2416&3.61&10.09&39.29&0.6867\\
10&0.2828&3.03&10.07&189.63&3.9565\\
11&0.2696&3.02&10.06&150.22&3.1478\\
12&0.2701&2.96&10.06&159.56&3.4113\\
13&0.3022&2.8&10.04&372.72&8.4406\\
14&0.2995&2.8&10.04&299.85&6.7904\\
15&0.2999&2.8&10.04&365.64&8.2802\\
16&0.3105&2.82&10.02&471.96&10.6333\\
17&0.3261&2.9&10.02&483.08&10.5836\\
18&0.3249&2.86&10.01&553.36&12.3052\\
19&0.3584&3.17&10.03&588.5&11.7833\\
20&0.357&3.2&10.03&606.16&12.0231\\
21&0.3592&3.15&10.01&610.296&12.3219\\
22&0.3626&3.19&10.02&616.08&12.2704\\
23&0.3661&3.24&10.02&620.97&12.1769\\
24&0.3775&3.28&10.01&645.44&12.5149\\
25&0.3726&3.28&10.02&620.53&12.0199\\
26&0.2716&3.58&10.08&59.25&1.0453\\
27&0.2693&3.56&10.07&78.82&1.3997\\
28&0.2659&3.05&10.06&110.58&2.2943\\
\hline
\hline
\label{tab:flat tablets}
\end{tabular}
}
\end{center}
\end{table}

\bibliographystyle{plainnat}

\end{document}